# Ga$_{Zn}$-V$_{Zn}$ acceptor complex defect in Ga-doped ZnO

*Aihua Tang[1,4], Zengxia Mei[1,*], Yaonan Hou[2], Lishu Liu[1], Vishnukanthan Venkatachalapathy[3], Alexander Azarov[3], Andrej Kuznetsov[3], and Xiaolong Du[1,4,*]*

[1]*Key Laboratory for Renewable Energy, Beijing National Laboratory for Condensed Matter Physics, Institute of Physics, Chinese Academy of Sciences, Beijing 100190, China*

[2]*Department of Electronic and Electrical Engineering, University of Sheffield, Broad Lane, S3 7HQ, Sheffield, UK*

[3]*Department of Physics, Centre for Materials Science and Nanotechnology, University of Oslo, PO Box 1048 Blindern, N-0316 Oslo, Norway*

[4]*School of Physical Sciences, University of Chinese Academy of Sciences, Beijing 100190, China*

*E-mails: zxmei@iphy.ac.cn; xldu@iphy.ac.cn*

Identification of complex defect has been a long-sought-after physics problem for controlling the defect population and engineering the useful properties in wide bandgap oxide semiconductors. Here we report a systematic study of (Ga$_{Zn}$-V$_{Zn}$)$^-$ acceptor complex defect via zinc self-diffusion in Ga-doped ZnO isotopic heterostructures, which were conceived and prepared with delicately controlled growth conditions. The secondary ion mass spectrometry and temperature-dependent Hall-effect measurements reveal that a high density of controllable (Ga$_{Zn}$-V$_{Zn}$)$^-$ is the predominant compensating defect in Ga-doped ZnO. The binding energy of this complex defect obtained from zinc self-diffusion experiments (~0.78 eV) well matches the electrical activation energy derived from the temperature-dependent electrical measurements (~0.82 ± 0.02 eV). The compensation ratios were quantitatively calculated by energetic analysis and scattering process to further validate the compensation effect of (Ga$_{Zn}$-V$_{Zn}$)$^-$ complex in Ga-doped ZnO. Meanwhile, its energy level structure was suggested based on the photoluminescence spectra, and the lifetime was achieved from the time-resolved photoluminescence measurements. The electron transitions between the (Ga$_{Zn}$-V$_{Zn}$)$^-$ complex defect levels emit the light at ~650 nm with a lifetime of 10-20 nanoseconds. These findings may greatly pave the way towards novel complex defects-derived optical applications.

# I. INTRODUCTION

It is generally accepted that point defects hold a major influence on the physical properties of advanced semiconductors, considering the fact that they dominate various diffusion mechanisms involved in doping and its limitation, processing and device degradation, minority carrier lifetime, and luminescence efficiency etc. [1–5]. Understanding of defects is a must for predictable defect engineering for useful properties. In spite of a significant research effort made worldwide, there is a lack of reliable experimental data on energetics of point defects in ZnO. Strikingly, +2 charged oxygen vacancy ($V_O$) has been experimentally recognized as the origin of n-type conductivity in ZnO from the energetic study of oxygen self-diffusion in isotopic heterostructures [6]. Zinc vacancy ($V_{Zn}$) energetics was further validated from zinc self-diffusion measurements [7]. Regarding more complicated extrinsic complex defects, it remains challenging to uncover their structural compositions and electrical features, not to mention the energetic characteristics. For example, $Al_{Zn}$–$V_{Zn}$ complex has been implied to be the main compensating defect in Al-doped ZnO via synchrotron x-ray absorption near edge structures and electron paramagnetic resonance spectroscopy techniques [8,9]. Generally, most researches of heavily n-doped ZnO are focused on promoting its electrical properties, recently on exploiting its exotic optical properties in surface and bulk plasmons [10,11]. However, a systematic study on identification of the extrinsic complex defects and their energetics has not yet been well established in terms of experimentation.

In this letter, secondary ion mass spectrometry (SIMS) combined with temperature-dependent Hall-effect (TDH) measurements was employed to investigate the electrically active $V_{Zn}$–related complex defect and its compensation behavior in Ga-doped ZnO isotopic heterostructures. SIMS in conjunction with isotope tracing atoms is a very powerful approach to studying point defects, because diffusion phenomenon is closely associated with defect formation process [6,7,12–14]. The Ga-doped ZnO isotopic heterostructures were conceived and prepared with delicately controlled growth conditions, such as chemical potential ($\mu$), Fermi level ($E_F$), and pre-annealing of the bottom layer in the reference sample. In this case, the energies derived from the analysis of self-diffusion can be distinctly interpreted and related to a specific point defect, due to their different dependence on the chemical potential and Fermi level [6,7,15–17]. A comparative study of the diffusion profiles reveals that $(Ga_{Zn}\text{-}V_{Zn})^-$ is the predominant compensating defect in Ga-doped ZnO, and the derived binding energy (~0.78 eV) agrees well with the electrical activation energy obtained from the TDH measurements. In addition, the energy level structure and carrier lifetime of $(Ga_{Zn}\text{-}V_{Zn})^-$ complex defect were also demonstrated via photoluminescence (PL) and time-resolved photoluminescence (TRPL) spectra.

# II. EXPERIMENTS

Two series of samples were in focus: ZnO isotopic double-layer heterostructures for self-diffusion study denoted with initial 'D', and ZnO single-layer films for TDH measurement denoted with initial 'S'. Two sorts of zinc sources were adopted in the synthesis of zinc-isotope Ga-doped ZnO heterostructure samples on c-oriented sapphire substrate by radio frequency plasma assisted molecular beam epitaxy (rf-MBE): one is in its natural isotopic ratio (labeled as $^n$Zn), and the other artificially enriched with 99.4% $^{64}$Zn (labeled as $^{64}$Zn). Ga-doped zinc-isotope ZnO heterostructures were synthesized with a top $^n$ZnO: Ga layer on a bottom $^{64}$ZnO: Ga layer, as schematically illustrated in the inset of Fig. 1(a). It should be noted that the $^{68}$Zn isotope atoms in the top $^n$ZnO layer were chosen as the diffusion source, and the bottom $^{64}$ZnO layer as the diffusion space. Reflection high-energy electron diffraction (RHEED) was applied in situ to monitor the whole epitaxial growth process (see Fig. S1 in the supplemental material). Diffusion anneals were performed in air for 2h, in a temperature range of 873 K - 973 K. Concentration depth profiles of Zn isotopes and dopants were characterized by SIMS with a Hiden MAXIM Analyser. The signal-to-concentration calibration was performed using standard $^n$Zn and $^{64}$Zn samples as a reference. The conversion of SIMS sputtering time-to-depth profiles was performed by measuring the crater depth using a Dektak 8 profilometer and assuming a constant erosion rate. TDH measurements were performed in an HL5500PC Hall Effect Measurement System. The transport properties of the samples were measured in a Quantum design PPMS machine. PL spectra were recorded at room temperature by exciting the samples with a 325 nm He-Cd laser with an output power of 136 μW. TRPL measurements were carried out with a 50 ps pulsed laser at an excitation wavelength of 375 nm. The time resolution of our system is less than 100 ps.

### III. RESULTS AND DISCUSSION

It is well known that the formation energy ($E^f$) of a point defect depends linearly on $\mu$ and $E_F$. The lower carrier concentration and mobility in O-rich Ga-doped ZnO [18,19] implies the existence of acceptor-like point defects with lower $E^f$ in O-rich conditions. One possible candidate, interstitial oxygen ($O_i$), can be reasonably excluded considering its relatively high formation energy and significant instability [15,20]. On the other hand, intrinsic $V_{Zn}$ and dopant-induced extrinsic $V_{Zn}$-related complex defects are still under debate. In present work, the chemical potential ($\mu$) and Fermi level ($E_F$) were delicately designed for Ga-doped ZnO samples, the amount and species of these $V_{Zn}$-related compensation defects in Ga-doped ZnO can be controlled to advantage the identification of their natures and energetic properties.

Note that the chemical potentials of $^{64}$Zn ($^n$Zn) and O, as well as the Ga dopant content, were kept the same for both the top and bottom ZnO: Ga layers in each sample, respectively. For different self-diffusion samples, $^{64}$Zn and $^n$Zn fluxes change while O flux keeps constant to make the concentrations of compensating defects in the samples vary with different $\mu$. $^{64}$Zn and $^n$Zn fluxes were the same for D1 and D2 (Zn rich), while decreased for D3 (O rich). On the other hand, D1 and D3 has a similar Ga concentration (~1×10$^{19}$ cm$^{-3}$, ~2×10$^{19}$ cm$^{-3}$, respectively), one order of magnitude lower than that of D2 (~1×10$^{20}$ cm$^{-3}$). Obviously, D1 is a reference sample with the least compensation defects, because it is prepared under an O-poor condition and with fewer Ga dopants incorporated [18,21,22]. In addition, aiming to dissociate the existing complex defects and supply ready $V_{Zn}$ sites for subsequent self-diffusion process, an in situ pre-annealing was deliberately carried out to the bottom $^{64}$ZnO: Ga layer before the deposition of top $^n$ZnO: Ga layer in D1 [21]. (See Table I for an overview of these samples). The Zn isotopic atom's self-diffusion profiles in these samples will reflect the dependence of $V_{Zn}$-related defects' formation on $\mu$ and $E_F$. Consequently, we can figure out its species and energetic properties.

The curves in Fig. 1(a) show the $^{68}$Zn concentrations versus depth profiles in as-grown D1-D3. Interestingly, the $^{68}$Zn tracing atoms in these samples manifest nearly identical profiles, suggesting the as-grown Zn diffusion-related defects almost not changing with varied $\mu$ or $E_F$, which will be discussed later. In addition, the Fermi levels $E_F(T)$ of D1-D3 were derived by applying TDH effect measurements in a temperature range of 523 K - 753 K, shown in Fig. 1(b). The $E_F$ levels of D1 and D3 almost keep the same as expected, while obviously lower than that of D2 at the corresponding temperatures.

As illustrated in Fig. 2(a), $^{68}$Zn atoms' diffusion in D1 becomes more pronounced at elevated annealing temperatures, which was the same for D2 and D3 (not shown here). Taking the curves annealed at 923 K as an example, the diffusion of $^{68}$Zn atoms was much more enhanced, in an order of D2, D3, and D1, as shown in Fig. 2(b). It indicates that the formation of dominant defects is more preferred under O-rich and high $E_F$ condition, as above mentioned. This finding is consistent with the broad consensus of $V_{Zn}$-mediated diffusion in ZnO [7,23,24]. The diffusion profiles were simulated using second Fick's law applying reflective boundary conditions with the as-grown profile as initial condition to obtain Zn self-diffusion coefficient, i.e. diffusivity D [6,7]. Based on the Arrhenius plots of D values versus the corresponding reciprocal absolute temperatures presented in Fig. 3(a), the diffusion activation enthalpy $\Delta H_a$ can be therefore extracted as 1.49 ±0.05 eV, 2.26 ±0.04 eV, and 2.28 ±0.03 eV for D1, D2, and D3, respectively.

It has been well established that $\Delta H_a$ of the diffusion defect can be expressed as the sum of the formation enthalpy $\Delta H_f$ and the migration enthalpy $\Delta H_m$, controlled by a thermally-activated intrinsic point defect mediated process. Under this assumption, for a certain point defect, $\Delta H_m$ will not be affected by $E_F$ and $\mu$, and $\Delta H_f$ is linearly related to $E_F$ and $\mu$. Specifically for $V_{Zn}$, $\Delta H_f$ and $\Delta H_a$ should decrease under more O-rich and higher $E_F$ conditions. The relatively low $\Delta H_a$ (1.49 ±0.05 eV) for D1 is basically equal to the theoretically predicted migration barrier (~1.4 eV) for $V_{Zn}$ [15,25]. It suggests that a direct diffusion mechanism without the need of forming thermally-activated defects (the diffusing atoms simply jump into neighboring vacancy sites) dominates the diffusion process as expected, i.e. the activation enthalpy of D1 is essentially determined by the migration enthalpy [6,26]. Intriguingly, $\Delta H_a$ for D2 (higher $E_F$) and D3 (more O-rich) is larger than D1. It differs from above-mentioned assumption and our previous results on Zn self-diffusion via thermally-activated intrinsic $V_{Zn}$ defect at relatively high-temperature region [7].

Firstly, the influence of dislocation motion and grain boundary diffusion can be excluded due to the similar crystal qualities for all samples (see Fig. S2 in the supplemental material). Moreover, their diffusion coefficients are a few orders of magnitude higher than that of volume diffusion, while their activation energies should be much lower than the values listed here [27,28].

Intuitively, the diffusion process should not be solely controlled by the intrinsic thermally-activated $V_{Zn}$. In fact, the regular curve of diffusion activation energy versus temperature can be divided into two parts: one of intrinsic control and one of extrinsic control [29]. Considering the anomalous finding that $\Delta H_a$ for D2 and D3 are obviously larger than D1, present diffusion process is reasonably attributed to the result of Ga doping-induced extrinsic defects rather than thermally-activated intrinsic $V_{Zn}$. As doping levels increase, the defect chemistry is completely dominated by extrinsic defects, and $\Delta H_a$ will reach a minimum that corresponds to $\Delta H_m$, the migration energy. Therefore, $\Delta H_a$ of D1 is almost identical to $\Delta H_m$ for intrinsic $V_{Zn}$.

In this case, if the samples are doped further, $\Delta H_a$ will remain fixed while $D_0$ will increase, because the concentration of mobile defects will increase along with the increased Ga dopant atoms. As illustrated in Fig. 2(b), $^{68}Zn$ self-diffusion in D2 becomes more pronounced than in D1 at the same temperature, which indicates larger quantities of extrinsic defects existing in D2 than D1. Excluding the influence of intrinsic $V_{Zn}$ introduced by thermal activation process, the enhanced self-diffusion behavior should only be caused by the dissociation of $V_{Zn}$-related complex (more accurately, $(Ga_{Zn}-V_{Zn})^-$ complex defect), which injects native defects during annealing [30]. Formation of $(Ga_{Zn}-V_{Zn})^-$ complex defects is much easier in D3 due to its more O-rich growth condition, and that's why the $^{68}Zn$ self-diffusion in D3 is also more enhanced than in D1 at the same temperature, as shown in Fig. 2(b). Furthermore, the almost identical $^{68}Zn$ profile curves in Fig. 1(a) also serves as solid evidence of this argument. The negligible influence of $\mu$ and $E_F$ on the $^{68}Zn$ concentration depth profiles of the as-grown samples indicates the available isolated Zn vacancies have ignorable effect on the diffusion of Zn atoms. Nevertheless, the diffusion profiles at 923 K manifest a strong dependence on $\mu$ and $E_F$. It suggests that the thermally dissociated Zn vacancies from $(Ga_{Zn}-V_{Zn})^-$ complex dominate the diffusion process. In other word, the $(Ga_{Zn}-V_{Zn})^-$ complex defect rather than isolated $V_{Zn}^{2-}$ should be the prevailing acceptor in Ga-doped ZnO. Since the $\Delta H_a$ of D1 is approximately equal to $\Delta H_m$ (an in situ pre-annealing was performed to the bottom $^{64}ZnO$: Ga layer of D1, aiming to dissociate complex defects and eliminate their influence on diffusion activation energy), the energy difference (~0.78 eV) between D2 (or D3) and D1 should be the activation enthalpy for the dissociation of $(Ga_{Zn}-V_{Zn})^-$ complex, also known as the binding energy. It is very close to the theoretical result (~0.75 eV) predicted by DFT calculations [21]. The binding energy of a complex is defined as the formation energy difference of a complex and its constituents. Theoretically, it is usually unaffected by $E_F$ and $\mu$ variations, which is quite consistent with our experimental findings.

The electrical compensation of $(Ga_{Zn}-V_{Zn})^-$ complex and its binding energy were further corroborated by the measurements of temperature-dependent carrier concentrations of S-series samples. Under a similar growth condition with D2, S1 was synthesized with a single Ga-doped ZnO epilayer structure. S2 was prepared in a similar way as S1, except a lower growth temperature (see the details in Table I). The electrical activation energy of S1 is determined by fitting the carrier concentration curve with the equation of n $\propto \exp(-E_a/k_B T)$, as demonstrated in Fig. 3(b). The value (~0.82 ±0.02 eV) well matches the binding energy of $(Ga_{Zn}-V_{Zn})^-$ complex defect inferred from zinc self-diffusion experiments (~0.78 eV), which further confirmed the electrical compensation in Ga-doped ZnO should be induced by the dominant $(Ga_{Zn}\text{-}V_{Zn})^-$ complex acceptor defect [21].

When coexisting in one matrix, $Ga_{Zn}$ and $V_{Zn}$ can form defect complexes via the following reactions:

$$Ga_{Zn}^+ + V_{Zn}^{2-} \rightarrow (Ga_{Zn} - V_{Zn})^- \quad (1)$$

$$Ga_{Zn}^+ + Ga_{Zn}^+ + V_{Zn}^{2-} \rightarrow (2Ga_{Zn} - V_{Zn})^0 \quad (2)$$

Formation of the three-defect complex, $(2Ga_{Zn}\text{-}V_{Zn})^0$, is statistically difficult, as its formation requires the diffusion of Ga impurities. However, since the Ga diffusion is much slower than the Zn self-diffusion because of the small impurity to host–cation ratio [31], it is assumed that the dissociation of $(Ga_{Zn}\text{-}V_{Zn})^-$ complex instead of $(2Ga_{Zn}\text{-}V_{Zn})^0$ is mainly involved in the annealing process. According to reaction (1), the associated detailed-balance relationship can be obtained as follows,

$$N_{Ga} \times N_V = N_{Ga-V} \times N_{site} \times e^{(-E_b/k_B T)} \quad (3)$$

Here, $N_{Ga}$ and $N_V$ are the concentrations of the isolated (unpaired) $Ga_{Zn}^+$ dopants and $V_{Zn}^{2-}$ defects, respectively. $N_{Ga-V}$ is the concentration of the $(Ga_{Zn}–V_{Zn})^-$ complex. $N_{site}$ is the number of sites that the defects can be incorporated in, $E_b$ is the binding energy of $(Ga_{Zn}–V_{Zn})^-$ complex, and $k_B$ is the Boltzmann constant. Feed $N_{site}$ (ZnO) = $4.28 \times 10^{22}$ cm$^{-3}$ and $E_b = 0.82$ eV into Equation (3) to obtain one relationship. In addition, considering the charge neutrality condition and the conservation of Ga atoms yields two additional relationships,

$$n = N_{Ga} - N_{Ga-V} - 2N_V, \quad (4)$$

$$N_{Ga}^{tot} = N_{Ga} + N_{Ga-V} \quad (5)$$

the values of $n = -9.88 \times 10^{17}$ cm$^{-3}$ and $N_{Ga}^{tot} = 2 \times 10^{20}$ cm$^{-3}$ were obtained from Hall and SIMS measurements, respectively, at room temperature. Solving Equations (3)-(5) for $N_{Ga}$, $N_V$ and $N_{Ga-V}$ yields $N_{Ga} = 1.00 \times 10^{20}$ cm$^{-3}$, $N_V = 8.52 \times 10^8$ cm$^{-3}$, and $N_{Ga-V} = 9.95 \times 10^{19}$ cm$^{-3}$. $N_{Ga-V} / N_V \approx 10^{11}$ indicates that nearly 100% of Zn vacancies are bounded in complexes. Therefore, the electrical compensation effect at room temperature mainly comes from $Ga_{Zn}\text{-}V_{Zn}$ complex rather than isolated Zn vacancy. The compensation ratio $K$ (i.e. $N_{Acceptor}/N_{Donor} = N_{Ga-V}/N_{Ga}$) can be thus deduced as high as 0.99. The small quantity of isolated Zn vacancies well interpret the reason why $\mu$ and $E_F$ have negligible influence on the $^{68}$Zn concentration depth profiles of the as-grown samples. However, after annealing at 1023 K in vacuum, n changes from $-9.88 \times 10^{17}$ cm$^{-3}$ to $-6.81 \times 10^{19}$ cm$^{-3}$. $N_{Ga} = 1.36 \times 10^{20}$ cm$^{-3}$, $N_V = 1.94 \times 10^{18}$ cm$^{-3}$, and $N_{Ga-V} = 6.40 \times 10^{19}$ cm$^{-3}$ can be obtained. Therefore, for annealed S1, $N_{Ga-V}/N_V$ reduces to 33, which means a large number of complex defects were dissociated through high-temperature annealing. The electrical compensation effect caused by $(Ga_{Zn}\text{-}V_{Zn})^-$ is remarkably weakened, and the compensation ratio $K$ is hence significantly reduced to 0.47.

The analysis of scattering process based on the mobility model for degenerate materials presented by Look et al. [32,33] further reveals the compensation effect of $(Ga_{Zn}\text{-}V_{Zn})^-$ complex defect (see the calculation details in the supplemental material). For annealed S1, the value of $K = 0.33$ is close to 0.47 obtained by energetic analysis hereinbefore. The acceptable difference between the two values may originate from the neglected thimbleful $(2Ga_{Zn}-V_{Zn})^0$ complex in energetic analysis. Besides the chemical potential, the growth temperature also significantly influences the amount of $(Ga_{Zn}-V_{Zn})^-$ complex defect. When the growth temperature was lowered to 723 K in S2, the carrier concentration and mobility are greatly enhanced compared to those of S1 (prepared at 873 K). The $K$ of S2 reduced to 0.12 combined with a dramatic decrease of $(Ga_{Zn}-V_{Zn})^-$ concentration. Overall, the results contribute to both an essential understanding and constructive engineering routes of $(Ga_{Zn}-V_{Zn})^-$ complex defect in heavily Ga-doped ZnO.

The room-temperature PL spectra of the as-grown D3, D3 after annealed at 1023 K in vacuum (D3 AN), and intrinsic ZnO (S3, which has the same $\mu$ as D3) are plotted in Fig. 4(a). All samples demonstrate two distinct emission bands: a signature ultraviolet (UV) near-band edge (NBE) emission of ZnO, and a quite broad visible emission band universally assigned to intrinsic or extrinsic defects, known as deep-level emission (DLE). The near-infrared (NIR) emission peak at ~760 nm is a second order of the UV NBE peak. Compared with D3 AN and S3, D3 shows a much weaker NBE emission and an obviously stronger red-shifted DLE peaked at ~650 nm. Generally speaking, $V_{Zn}$-related acceptor defects were assumed to be responsible for the emission near 650 nm [34,35]. In present study, we believe it belonging to the two-level emitter of $(Ga_{Zn}-V_{Zn})^-$ instead of single $V_{Zn}$, as schematically illustrated in the inset of Fig. 4(a). According to the above-mentioned data, $(Ga_{Zn}-V_{Zn})^-$ complex defect rather than the isolated Zn vacancy acts as the main acceptor in Ga-doped ZnO and will dissociate significantly after annealing. Dissociation of substantial $(Ga_{Zn}-V_{Zn})^-$ complex defects causes a remarkable blue shift of the DLE toward the widely studied "green" band near 550 nm. Apparently, the high density of ionized $(Ga_{Zn}-V_{Zn})^-$ complex defect, contributing to the broad DLE emission at ~650 nm, results in an enhanced nonradiative transition and largely reduced NBE [36,37]. Note that, due to the heavily electrical compensation in our as-grown samples, the concentrations of free carriers in them are very low (~$10^{17}$ cm$^{-3}$), which is obviously lower than theoretically calculated critical concentration of degenerate state for Ga-doped ZnO (>$10^{18}$ cm$^{-3}$). Moreover, the Fermi levels shown in Fig. 1(b) also indicate all as-grown Ga-doped samples are non-degenerate, so the upper level of the $(Ga_{Zn}-V_{Zn})^-$ complex defect will not merge into the conduction band. A nonradiative transition firstly happens to the photo-generated electrons from the conduction band to the upper level of the $(Ga_{Zn}-V_{Zn})^-$ complex defect, which should be close to the conduction band minimum. Then a large number of electrons transits between the two states, causing the intense "red" DLE. The shallow donor states of isolated $Ga_{Zn}^+$ are activated by dissociating the $(Ga_{Zn}-V_{Zn})^-$ complex, which leads to the band-gap renormalization effect and hence the red shift of NBE [36].

The photodynamic characteristics of $(Ga_{Zn}-V_{Zn})^-$ acceptor complex defects was further explored by TRPL spectra. Fig. 4(b) shows the decay profiles of D3, D3 AN and S3. They were fitted by a triple-exponential function with time constants $\tau_1$, $\tau_2$, and $\tau_3$. The fast decay component $\tau_1$ is commonly attributed to nonradiative recombination [38], and the slow decay component $\tau_3$ is ascribed to electron communications with the background, which is usually a complex process with a long relaxation time. The remaining component $\tau_2$ (16.46 ns, 2.75 ns, and 0.78 ns for D3, D3 AN and S3, respectively) presents the lifetime of deep level defects. As demonstrated in Fig. 4(b), the decay profile of visible emission peak of D3 (~650 nm) is obviously different from those of D3 AN and S3 (~550 nm). The lifetime of $(Ga_{Zn}-V_{Zn})^-$ complex defect emission is quite larger than that of "green" emission (so are other Ga-doped ZnO samples, see Fig. S7 and Fig. S8 in the supplemental material). It implies that the involved emission mechanisms are different. As demonstrated in Fig. 4(a), the DLE of D3 AN behaves the same as S3. The "green" band emission of D3 AN and S3 is generally ascribed to the transition from the conduction band to deep defect level, which was attributed to single positively charged oxygen vacancy ($V_O^+$) in some previous work [35,39,40]. Therefore, the initial state of $(Ga_{Zn}-V_{Zn})^-$ complex defect emission is not the conduction band, further confirming the two-level emitter model of $(Ga_{Zn}-V_{Zn})^-$ complex defect. Consequently, the recombination of electrons between the $(Ga_{Zn}-V_{Zn})^-$ complex defect levels emits the light centered at 650 nm with a lifetime of 16.46 nanoseconds, which is closed to the reported lifetime of single photon emission (SPE) in ZnO film [41].

A renewed interest in ZnO has been fueled by its attractive prospects as the host of photonic qubits at room temperature [42]. Actually, point defect-derived optical applications of ZnO have gained momentum in recent years, since a room-temperature SPE being continually reported [41,43–48]. Constrained by the lack of reliable experimental data on the point defect energetics in ZnO, the origin of these luminescent centers is still unclear although it was supposed to be related with zinc vacancy [41,46–48]. Moreover, engineering of two-state emitter defects is necessarily important for device applications. The investigating the photoluminescence characteristics of $(Ga_{Zn}-V_{Zn})^-$ complex defect may shed light on this research area. On the other hand, the study of defect-surface plasmon coupling in this heavily compensated ZnO: Ga material has been systematically carried out by optical characterizations [49], based on the achievement of a high density of controllable $(Ga_{Zn}-V_{Zn})^-$ in Ga-doped ZnO. Therefore, the distinct identification of $(Ga_{Zn}-V_{Zn})^-$ complex defect and its optical properties will advance the exploration of its new applications in single photon sources and high-efficiency nanophotonic emitters.

## IV. CONCLUSION

In conclusion, investigation of Zn self-diffusion behaviors in ZnO: Ga isotopic heterostructures with varied $\mu$ and $E_F$ yields the unambiguous identification of $(Ga_{Zn}-V_{Zn})^-$ acceptor complex defect as the predominant compensating center. Understanding of its energetic properties is successfully achieved by the combined SIMS and TDH measurements on deliberately designed samples. The experimental binding energy of ~0.78 eV agrees well with the theoretical result. A self-consistent electrical activation energy value of 0.82 eV $\pm$ 0.02 eV is determined via the TDH measurements, further confirming the prevailing role of $(Ga_{Zn}-V_{Zn})^-$ in compensating the carriers in Ga-doped ZnO. The compensation ratio $K$ ($N_{Ga-V}/N_{Ga}$) in Ga-doped ZnO is calculated and corroborated by energetics analysis and scattering process, respectively. Moreover, the energy level structure and lifetime of the $(Ga_{Zn}-V_{Zn})^-$ complex defect were revealed via PL and TRPL technique. The recombination of electrons between the $(Ga_{Zn}-V_{Zn})^-$ complex defect levels emits the light at ~650 nm with a lifetime of 10~20 nanoseconds.

Altogether, this work presents a unique isotope tracing solution to studying complex point defects and their energetic properties, and may greatly pave the way towards novel complex defect-derived optical applications.

## ACKNOWLEDGMENTS

This work was supported by the National Natural Science Foundation of China (Grants No. 11674405, 11675280, 11274366, 51272280, and 61306011).

Table I. The growth conditions and electrical properties of D- and S-series samples

| Sample name | Substrate temperature (℃) | Chemical potential | Ga concentration (cm$^{-3}$) | Thickness (nm) | Carrier concentration (cm$^{-3}$) | Mobility (cm$^2$V$^{-1}$s$^{-1}$) |
|---|---|---|---|---|---|---|
| D1 | 600 | Zn rich | ~1×10$^{19}$ | 620 | -4.96×10$^{17}$ * | 20.7 * |
| D2 | 600 | Zn rich | ~1×10$^{20}$ | 680 | -7.37×10$^{17}$ | 24.2 |
| D3 | 600 | O rich | ~2×10$^{19}$ | 670 | -4.80×10$^{17}$ | 17.8 |
| S1 | 600 | Zn rich | ~2×10$^{20}$ | 350 | -9.88×10$^{17}$ | 17.5 |
| S1 1023 K AN | -- | -- | -- | -- | -6.81×10$^{19}$ | 55.2 |
| S2 | 450 | Zn rich | ~3×10$^{20}$ | 420 | -1.24×10$^{20}$ | 69.8 |
| S3 | 600 | O rich | -- | 250 | -1.08×10$^{16}$ | 66.0 |

* An in-situ pre-anneal treatment was performed to the bottom $^{64}$ZnO: Ga layer of D1, to further decrease the compensation defects and provide V$_{Zn}$ sites available in the bottom diffusion space layer.

**Figure Caption:**

**FIG. 1.** (a) $^{68}$Zn concentration versus depth profiles in as-grown D1-D3. A schematic structure of these isotopic samples is shown in the inset. (b) Fermi level and carrier concentration versus temperature in the range of 523 K-753 K. The dash line indicates the band gap.

**FIG. 2.** Diffusion profiles of $^{68}$Zn concentrations in D1-D3: D1 after annealing in air for 2h (a) and D1-D3 after annealing at 923 K in air for 2h (b).

**FIG. 3.** (a) Arrhenius plots of the extracted Zn self-diffusion coefficient *D* versus the reciprocal temperature 1000/T. The solid lines show the best fits to the self-diffusion coefficients. The inset manifests the obtained activation energies for D1-D3. (b) Temperature dependence of the carrier concentration (n~T) over a temperature range of 593 K-773 K for S1. The Arrhenius plot for the temperature range of 653 K-773 K and the linear fitting are drawn in the inset.

**FIG. 4.** (a) Room-Temperature PL spectra of D3, D3 after annealing (D3 AN) and intrinsic ZnO S3, which has the same chemical potential with D3. An energy level diagram for DLE is shown in the inset. (b) The delay profiles and their fitting curves of the center wavelength of the visible emission peaks of D3 (~650 nm), D3 AN (~550 nm) and S3 (~550 nm), respectively.

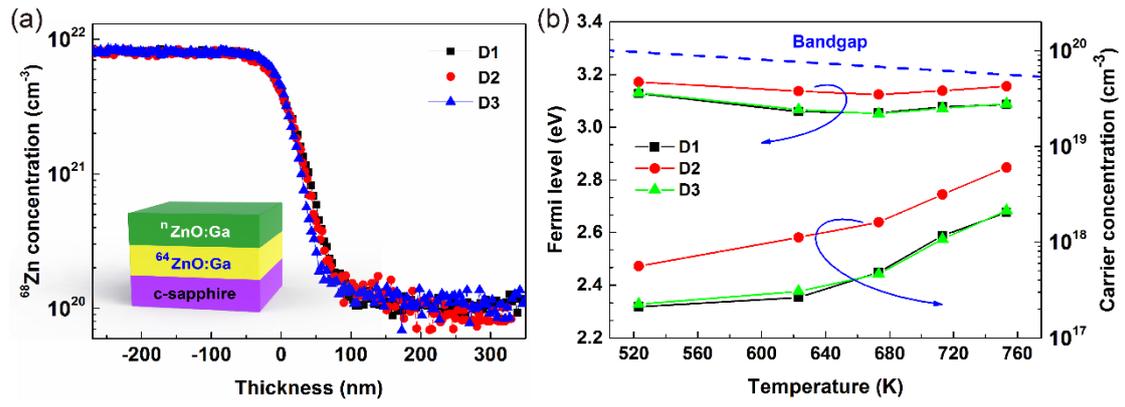

**FIG. 1**

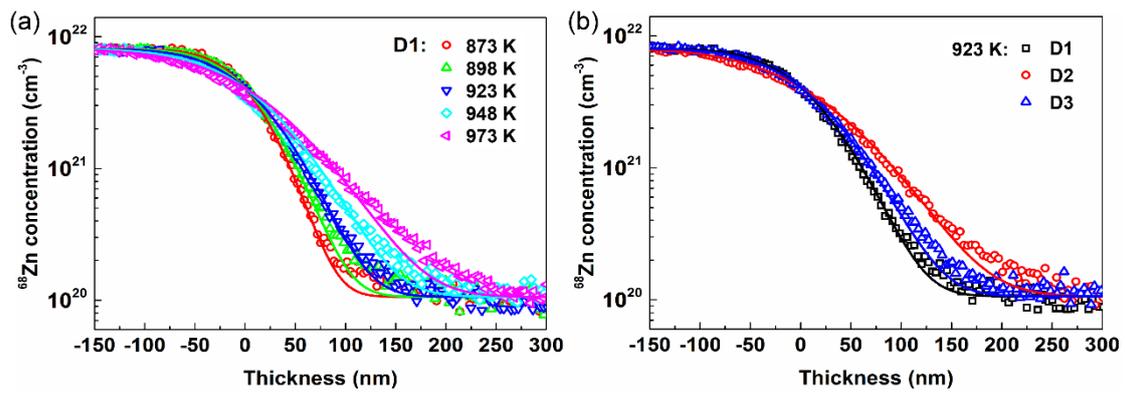

**FIG. 2**

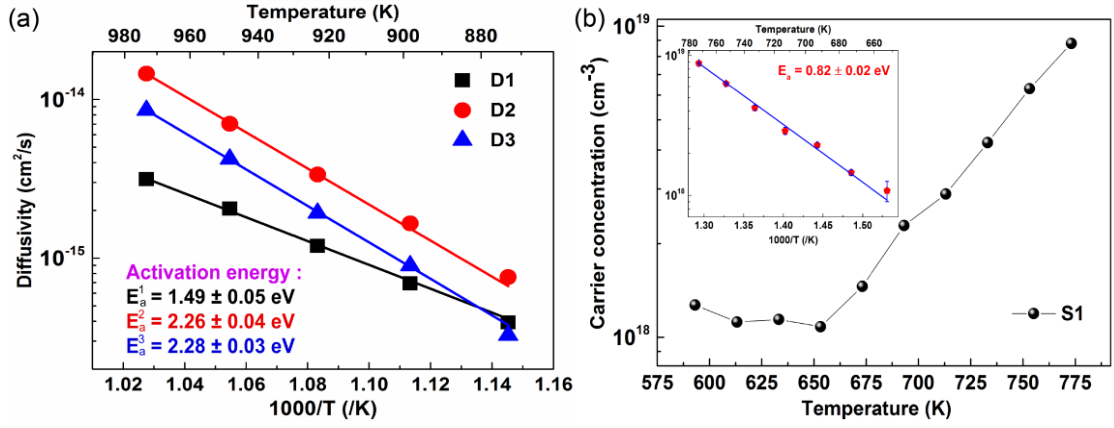

**FIG. 3**

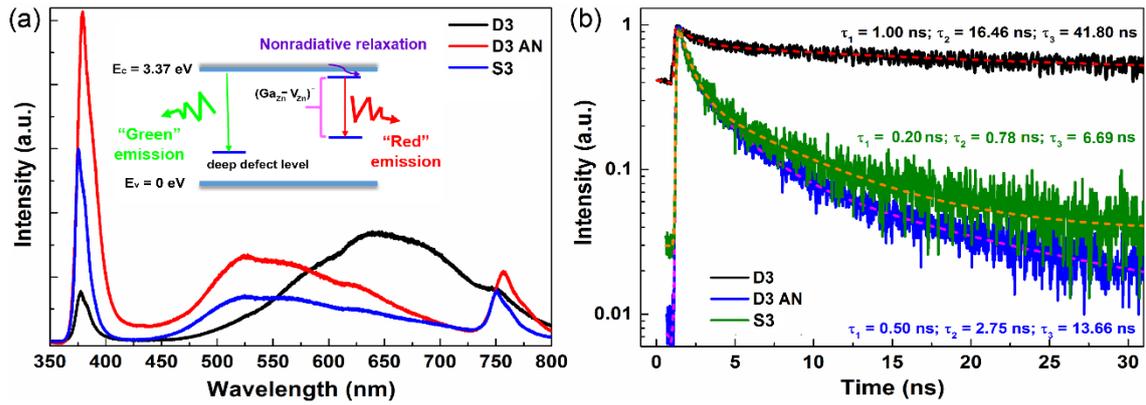

**FIG. 4**

**Additional Information**

Supplementary information is available from the author.